\documentclass[]{aastex}
\usepackage{psfig}
\def\gsim{\lower 2pt \hbox{$\, \buildrel {\scriptstyle >}\over
{\scriptstyle \sim}\,$}}
\def\lsim{\lower 2pt \hbox{$\, \buildrel {\scriptstyle <}\over
{\scriptstyle \sim}\,$}}

\def\rosat{{\sl ROSAT}}
\def\ROSAT{{\sl ROSAT}}
\def\asca{{\sl ASCA}}
\def\ASCA{{\sl ASCA}}
\def\ha{H$\alpha$}
\def\HI{H{\small I}\ }

\shortauthors{Wang}
\shorttitle{X-9 and its environs}

\begin{document}

\title{Ultraluminous X-ray Source 1E 0953.8+6918 (M81 X-9):\\ 
An Intermediate Mass Black Hole Candidate and its Environs
}
\author{Q. Daniel Wang}
\affil{Astronomy Department, University of Massachusetts, Amherst, MA 01003, USA}
\affil{Email: wqd@astro.umass.edu}

\begin{abstract}

We present a {\sl ROSAT} and {\sl ASCA} study of the {\sl Einstein} source X-9 
and its relation to a shock-heated shell-like optical nebula in a
tidal arm of the M81 group of interacting galaxies. Our {\sl ASCA} observation 
of the source shows a flat and featureless X-ray spectrum well 
described by a multi-color disk blackbody model. The source most likely 
represents an optically thick accretion disk around an intermediate mass 
black hole ($M \sim 10^2 {\rm M}_{\odot}$) in its high/soft state, similar to
other variable ultraluminous X-ray sources observed in nearby disk galaxies. 
Using constraints derived from both the innermost stable orbit around a
 black hole and the Eddington luminosity, we 
find that the black hole is fast-rotating and that its mass is between 
$\sim 20/({\rm cos}\ i)\ {\rm M}_{\odot} - 110/({\rm cos}\ i)^{1/2}
{\rm M}_{\odot}$, where $i$ is the inclination angle of the disk. The inferred
bolometric luminosity of the accretion disk is $\sim (8 \times 10^{39} 
{\rm~ergs~s^{-1}})/({\rm cos}\ i)^{1/2}$. Furthermore, we
find that the optical nebula is very energetic and may contain
large amounts of hot gas, accounting for a soft X-ray component as indicated
by archival \rosat\ PSPC data. The nebula is apparently associated with 
X-9; the latter may be powering the 
former and/or they could be formed in the same event (e.g., a hypernova). 
Such a connection, 
if confirmed, could have strong implications for understanding both the 
birth of intermediate mass black holes and the formation of energetic 
interstellar structures.
\end{abstract}


\section{Introduction}

       One of the most enigmatic X-ray-emitting objects is the source X-9 
(1E 0953.8+6918), discovered with the {\it Einstein 
Observatory} in the field close to the galaxy M81 (Fig.\ 1; Fabbiano 1988). 
The source is located about 12\farcm5 from the nucleus of the
galaxy and 2$^\prime$ from the galaxy's dwarf companion Ho IX 
($1^\prime$ corresponds to a projected separation of 1 kpc at the distance 
$D = 3.6$~Mpc; Freedman et al. 1994). 
The 0.2--4~keV flux of the source is only about a factor of $\sim 2$ lower 
than the flux of the nucleus (a LINER). X-9 was also detected in 
subsequent \rosat\ and \asca\ observations at 
comparable flux levels. Clearly the source is not a transient. 
X-9 does, however, exhibit significant sporadic timing 
variability and therefore must primarily be a compact
source (Immler \& Wang 2001; Ezoe et al. 2001).
Interestingly, Miller (1995) discovered that the source was projected inside 
a very  unusual \ha-emitting nebula (Fig.\ 1), which also emits 
strong [S{\small II}] and [O{\small I}] lines. He concluded that this 
nebula was shock-heated and might be a very energetic supernova remnant (SNR) 
or a superbubble. Furthermore, the nebula is apparently located 
within a nearly ``quiescent'' massive atomic and molecular 
gas concentration in a tidal arm (Concentration I, Yun et al. 1994; 
Fig.\ 1a). The lack of a significant stellar population in this region has
further led to the suggestion of the
concentration being a protogalaxy (Henkel et al. 1993). The nature of 
this combination of stellar and interstellar features remains
unknown.

\begin{figure}
\centerline{\hfil\hfil
\psfig{figure=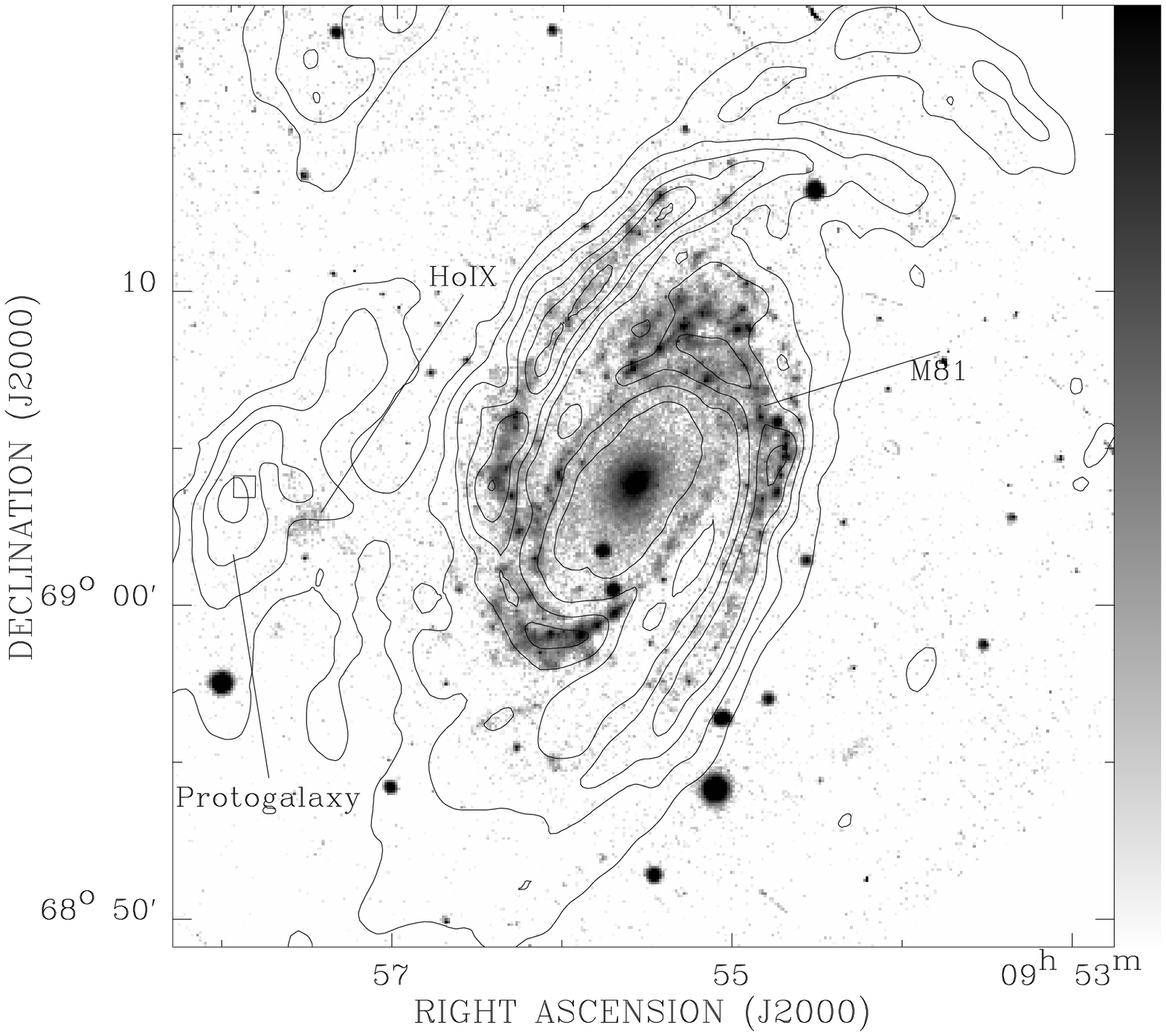,height=3.4truein,angle=90.0,clip=}
\hfil\hfil}
\centerline{\hfil\hfil
\psfig{figure=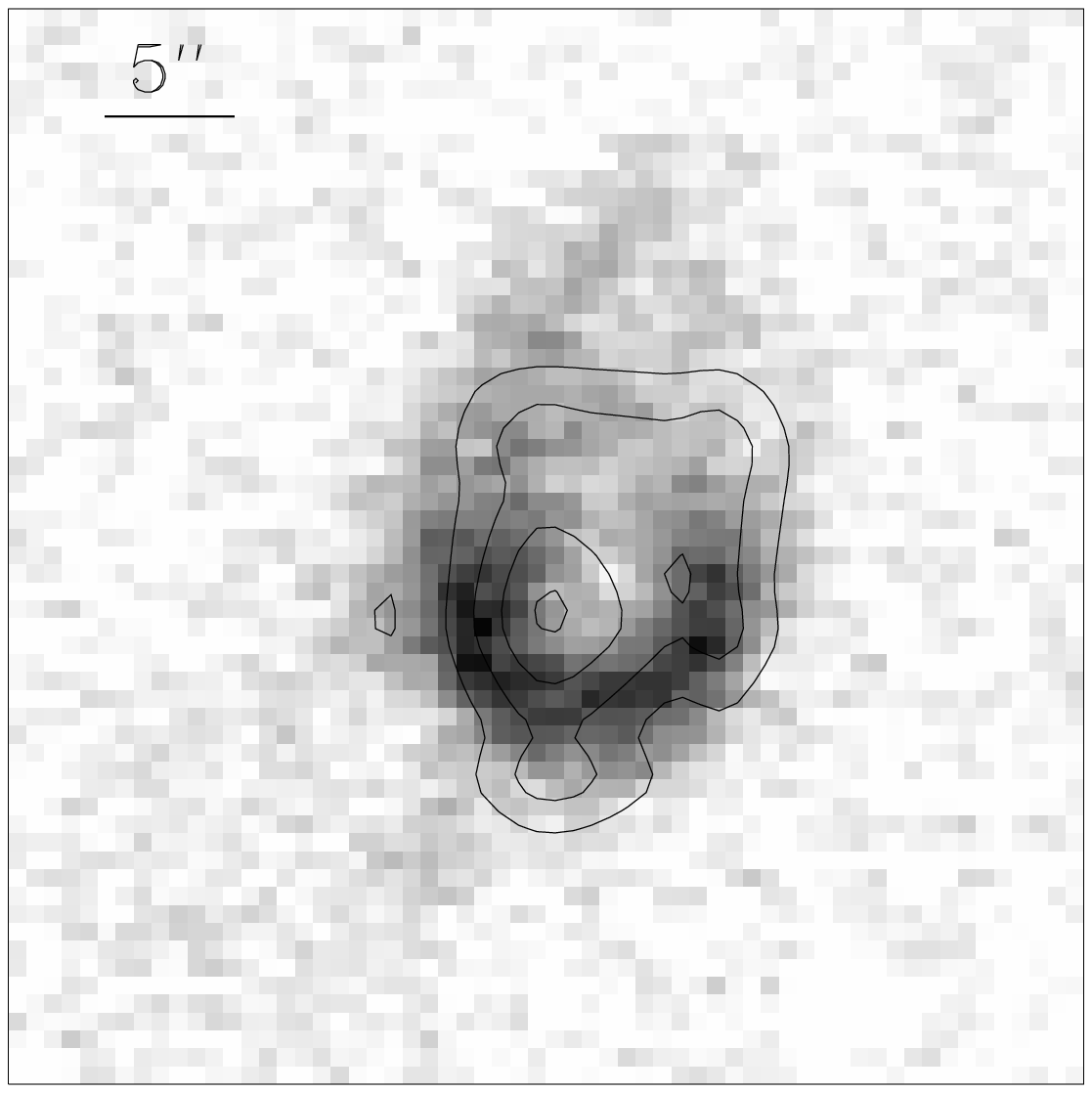,height=3.2truein,angle=0.0
 ,bbllx=152bp,bblly=395bp,bburx=474bp,bbury=715bp,clip=}
\hfil\hfil}
\caption{\protect\footnotesize  
The field including M81 in near-UV
($\sim 2490 $\.A; Hill et al. 1992), with overlaid HI contours 
at 5, 10, 15, 20, and 25 $\times 10^{20} {\rm~cm^{-2}}$ (upper panel). 
A close-up of the region outlined by a box in the 
field of the protogalaxy is shown in the lower panel. The 
H$\alpha$ image is provided by Miller (1995) and
the \rosat\ HRI intensity contours are plotted at 0.24, 0.34, 0.44, and 
0.54 $\times  {\rm~counts~s^{-1}~arcmin^{-2}}$. The structure of the X-ray
emission is uncertain because of the complicated PSF of the observations at
the large off-axis angle  (Immler \& Wang 2001). 
}
\end{figure}

	In this paper we present results from the analysis of a dedicated 
\ASCA\  and archival \ROSAT\ observations of the intriguing X-ray source X-9.
We concentrate on the nature of the source and its relation to the environs, not on the detailed processes involved in the emission and evolution of the source. 

\section{X-ray Observations}

Our study is based primarily on our \asca\ observation of X-9,
taken in April of 1999.  Located at ${\rm R.A., 
Dec.}$ $({\rm J2000})$ $=9^{\rm h}57^{\rm m}54^{\rm s}$, $69^\circ 03^\prime 50^{\prime\prime}$ 
(Source \# H44/P66; Immler \& Wang 2001), the source was 
observed with the two pairs of 
the SIS (exposure 41 ks) and GIS detectors (35 ks). 
The SIS, however, experienced a loss 
of sensitivity to low-energy ($\lesssim 2$~keV) photons 
since approximately late 1994. This 
loss increased with time and reached as much as 40\% at the time
of our observation. No method has been proposed to 
quantify this loss. Nevertheless, the data from the SIS, with 
a better spectral resolution than the GIS, show no 
sign of any emission line in the spectrum of X-9. The present 
study therefore includes only the data from the GIS, which 
had a spectral resolution of $\delta E/E \approx 0.08$$ (6 {\rm~keV}/E)^{0.5}$
over the 0.5--10 keV band and a 50\% half power diameter of $\sim 3^\prime$ on axis. 

	We followed the standard
data processing procedure as detailed by Day et al. (1995), using the HEAsoft 
software package. We extracted on-source counts from an aperture of 4\farcm5 
radius and estimated the background contribution in a circle of $15^\prime$ 
radius around the GIS image center, excluding regions of $6^\prime$ radii
around the M81 nucleus and X-9. The data from the two GIS detectors 
were first combined and then loaded into the  software XSPEC for
spectral analysis.

	We also incorporated archival data from \rosat\ PSPC and
\asca\ GIS observations in our analysis. All these observations were
pointed at either SN1993J or M81 and covered a period of about 9 years;
X-9 was included in the field of view at large off-axis angles 
($\gtrsim 10^\prime$) in each of the observations. 
The archival data were  partly analyzed by Ezoe et al. (2001) and
by Immler \& Wang (2001), concentrating on timing variability. We extracted
the archival GIS data in a similar fashion as described 
above for our observation. We combined PSPC spectra extracted 
from an on-source aperture of 1\farcm3 radius and a background annulus 
of inner and outer radii, 1\farcm8 and $5^\prime$, respectively. All the 
spectra are corrected for off-axis PSF and effective area effects.

\section{X-ray Data Analysis and Results}

Because X-9 is variable  and lacks an apparent point-like optical 
counterpart, we concentrate on spectral models that are appropriate for 
a compact X-ray source. We first fit the spectrum 
from our own \asca\ observation alone.
Table 1 summarizes results from the fits of several commonly used models. 
The power law is characteristic of AGN spectra in the \asca\ band, 
although they sometimes exhibit soft {\sl excesses} at energies 
$\lesssim 1$ keV and/or hard X-ray {bumps} at $\gtrsim 10$ keV. 
Previous analysis by Ezoe et al. (2001) showed that a power law 
gave an acceptable fit to an average X-ray spectrum of X-9, extracted from 
archival GIS observations. The quality of this spectrum is problematic, 
however, chiefly because of the poor PSF at off-axis angles greater than 
10$^\prime$. There might also be substantial uncertainties in background 
subtraction and energy-dependent PSF-related correction as well as in the 
calibration of the instrument energy response. Indeed, the power 
law model fits our high quality GIS spectrum of X-9 poorly.  The spectrum is 
convex-shaped with flux {\sl deficits} at both low and 
high energy ends and shows no sign of an Fe K line at $\sim 6.4 - 6.7$ keV, 
which often appears in the spectra of bright AGNs. Therefore, X-9 is unlikely 
to be an AGN. 

	Could X-9 be Galactic in origin?
We considered the black body, which is a reasonably good 
description of X-ray emission from the surface of an isolated neutron star. 
The model, however, is much too sharply peaked to 
satisfactorily represent the spectrum of X-9. Raymond \& Smith 
optically thin thermal plasma also gives an unacceptable fit 
to the spectrum. Our \asca\
observation with a timing resolution up to $\sim 0.1$~ms 
was designed partly to test the hypothesis that the source might be
a moving X-ray pulsar, which could also power the optical nebula. 
Our FFT and period folding timing analyses of 
X-9, however, show no significant periodic signal in the
period range from 1 ms to several hours. 

\begin{deluxetable}{lcc}
\tablecaption{Model fits to the \asca\ GIS spectrum}
\tablewidth{0pt}
\tablehead{
\colhead{Model} & \colhead{Parameter\tablenotemark{a}}& \colhead{$\chi^2/n.d.f.$}
}
\startdata 
Power-law & energy slope $=1.3(1.2-1.4)$& 240/140\\
& $N_{\rm H} = 5.0(4.5-5.6) \times 10^{21} {\rm~cm^{-2}}$\\
Blackbody & Temperature $=0.74$~keV & 511/140\\
& $N_{\rm H} = 2.3(0.9-3.2) \times 10^{21} {\rm~cm^{-2}}$\\
MCD & $T_{\rm in}=1.34(1.31-1.37)$~keV & 93/130 \\
& $K_{\rm MCD} = 0.20(0.18-0.22)$\\
& $L_{\rm x} = (1 \times 10^{40}/{\rm cos}\ i) {\rm~ergs~s^{-1}}$\\
& $N_{\rm H} = 3.0(0.0-6.1) \times 10^{20} {\rm~cm^{-2}}$\\
\enddata
\tablenotetext{a}{$T_{\rm in}$ is the characteristic temperature of the disk
at its inner radius. The normalization of the MCD model
is defined as $K_{\rm MCD} = (R_{\rm in}/D)^2 {\rm~cos}\ i$, where $R_{\rm in}$ is the
inner disk radius in units of km, the distance to the source $D$ is 
in units of 10 kpc,  and $i$ is the disk inclination angle. 
The disk luminosity $L_{\rm x}$ is calculated in the 0.5--10~keV 
band. 
Uncertainties in the parameter estimates (enclosed in parentheses) 
are at the 90\% confidence limits.}
\end{deluxetable}

	We find that an accreting black hole model is an attractive
option for the X-9 spectrum (Fig.\ 2). The source, 
if at the distance of M81 and radiating isotropically, has an X-ray 
luminosity of $\sim 1 \times 10^{40} {\rm~ergs~s^{-1}}$. This luminosity 
is a factor of $\sim 10^2$ greater than the Eddington luminosity
for a $1.4~{\rm M}_{\odot}$ accreting neutron star, but is within the range 
of the so-called ultra-luminous X-ray sources (ULXSs), which are detected 
typically in disk galaxies and considered as
accreting BHs in the high/soft state (Colbert, \& Mushotzky 1999; 
Makishima et al. 2000). 
The X-ray spectra of these sources are well described by the multi-color disk 
blackbody (MCD) model (XSPEC model {\sl diskbb}). Indeed, the model fits 
the spectrum of X-9 well (Table 1).

\begin{figure} 
\centerline{ {\hfil\hfil
\psfig{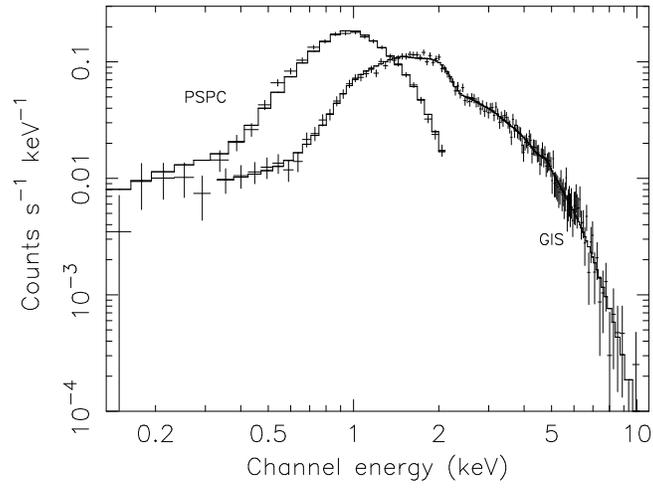}
\hfil\hfil}}
\caption{Joint fit to the combined \asca\ GIS and \rosat\ PSPC spectra 
of X-9. The histograms represent the best fit of the MCD model.
\label{fig2}}
\end{figure}

	Next, we jointly fit the GIS and PSPC spectra with the 
MCD model (Fig.\ 2). The PSPC data are particularly 
sensitive to soft X-ray emission and 
absorption. Because the flux of X-9 varies with time 
(amplitude of variability exceeding a factor of $\sim2.5$; cf. Fig.\ 3 in 
Immler \& Wang 2001) and because the spectrum could also 
change accordingly, we allow both $T_{\rm in}$ and $K_{\rm MCD}$ to be 
different for the GIS and PSPC spectra and joint-fit the X-ray absorbing
gas column density.
The MCD model gives a good fit to the spectra ($\chi^2/n.d.f. = 143/166$). 
The fitted column density is $1.0 (0.9-1.1) \times 10^{21} {\rm~cm^{-2}}$
 (90\% confidence limits), compared to the expected Galactic contribution
of $\sim 5 \times 10^{20} {\rm~cm^{-2}}$ in the direction. This 
extra absorption is consistent with the hypothesis that X-9 is located 
inside Concentration I. The MCD spectral parameters for the GIS data 
are nearly identical to those from the above single spectrum fit.
But the spectral parameters for the PSPC data
are rather different: $T_{\rm in} = 0.59(0.55-0.63)$ and $K_{\rm MCD} = 
1.7(1.3-2.2)$. 

\section{Discussion}

	 The above spectral characteristics of X-9, together with its
strong aperiodic variation (Immler \& Wang 2001) and its lack of 
periodicity, argues against a scenario of the source as
a Galactic compact object.
Indeed, at the source's Galactic position of $l, b = 143^\circ, 18^\circ$, 
there is little room for such a scenario.
Ezoe et al. (2001) have made additional arguments against the Galactic origin 
of X-9, based primarily on its high X-ray to optical emission ratio.
In contrast, the model of X-9 as an accreting BH inside Concentration I 
at the M81 distance naturally accounts for the sporadic timing behavior, 
the convex-shaped spectrum, the excess absorption, and the large X-ray flux
of the source. 

	But the apparent discrepancy between
the GIS and PSPC spectral characteristics demands an explanation. 
The low $T_{\rm in}$ as inferred from the PSPC spectrum is very abnormal,
compared with similar ULXSs observed with \asca. X-9 also shows
strong flux variation: e.g., $2.5 \times 10^{-12} 
{\rm~ergs~s^{-1}~cm^{-2}}$ for the PSPC spectrum vs. 
$5.1 \times 10^{-12} {\rm~ergs~s^{-1}~cm^{-2}}$ for the GIS spectrum
over the overlapping 0.5--2 keV band. But this variation is 
typical for ULXSs and is well within the range inferred from the 
archival GIS spectra of X-9. Indeed, the MCD model 
provides satisfactory fits to all individual GIS spectra, and
$T_{\rm in}$ is always greater than 1.2 keV. Clearly, the MCD model with the 
abnormal low $T_{\rm in}$ cannot be a fair 
characterization of the GIS spectra. It is important to note that 
the MCD model has been tested primarily on X-ray spectra of heavily 
absorbed Galactic black hole candidates. It is possible 
that the model does not adequately describe the soft X-ray emission 
from the accretion disk of X-9.

	Alternatively, the PSPC spectrum of X-9 may contain
a significant soft X-ray contribution from diffuse hot gas. This may be 
expected in the region enclosed by the shock-heated optical nebula, or 
from relativistic particles accelerated in an expanding shock
related to the nebula. The narrow bandwidth and very limited spectral 
resolution of the PSPC data, however, do not allow for a useful spectral
decomposition of such a soft component from the uncertain
disk emission of the source at the time of the observations.
A spatial analysis of the archival \rosat\ PSPC and HRI data also shows 
evidence for the extended X-ray emission around the centroid of X-9 on 
scales comparable to the size of the nebula. But
various systematic uncertainties (e.g., the source centroid shift 
caused by the off-axis instrumental point spread function)
prevent us from a definitive measurement of the extended emission. 

\subsection{X-9 as an Accretting IMBH}

Assuming that the MCD model is correct, we can estimate the mass 
of the BH: $M_{\rm BH} = (41 {\rm M}_{\odot}) \alpha^{-1} (K_{\rm MCD}
/{\rm cos}~i)^{1/2},$ where 
$\alpha$ is the ratio of the inner disk radius to the last stable
orbit radius of a non-rotating (Schwarzschild) BH, and 
both $K_{\rm MCD}$ and $i$ are defined in the note to Table 1. Adopting $K_{\rm MCD}$ 
in Table 1 gives 
$M_{\rm BH} =  18/(\alpha^2 {\rm cos}~i)^{1/2} {\rm M}_{\odot}$. 
For a general rotating Kerr BH, $\alpha \gtrsim 1/6$ 
(e.g., Zhang et al. 1997) and thus $M_{\rm BH} \lesssim 108 
{\rm M}_{\odot}/({\rm cos}~i)^{1/2}$. The model, however, does not include various
general relativistic effects (e.g., light bending). Ongoing modeling
of such effects shows that the cosine law is
approximately preserved for Schwarzschil BHs and that for 
extreme Kerr BHs the effective ${\rm cos}\ i = 0.17-0.4$ for $i = 
5^\circ-85^\circ$ (Zhang et al. 2001). So the range of variation can be
considerably smaller than the cosine law.
Following Makishima et al. (2000), we can also estimate the bolometric 
luminosity of the disk as
$L_{bol} = (8 \times 10^{39} {\rm~ergs~s^{-1}})/({\rm cos}~i)^{1/2}.$ 
Because $L_{bol}$ should typically be smaller than, or at most comparable 
to, the Eddington luminosity  $1.3 \times 10^{38} 
(M_{\rm BH}/{\rm M}_{\odot})$, 
we obtain $\alpha \lesssim 0.76 ({\rm cos}~i)^{1/2}$, 
indicating that the BH is fast-rotating ($\alpha < 1$) and that 
$M_{\rm BH} \gtrsim 24 {\rm M}_{\odot}/{\rm cos}\ i$. The 90\% statistical 
uncertainties in these mass limits are about 10\%.

\subsection{Optical Nebula}

	The presence of the unusual shell-like optical nebula 
is an important part of the mystery about X-9 (Fig.\ 1b). 
Optical spectroscopy of the nebula shows that its mean heliocentric
velocity ($\sim 47-52 {\rm~km~s^{-1}}$) agrees with the velocity
($\sim 45-65 {\rm~km~s^{-1}}$) of the \HI concentration 
(Adler \& Westpfahl 1996; Miller 1995). The \HI velocity field further 
shows that the concentration is part of the M81 group (Miller 1995). 
In contrast, the dwarf galaxy Ho IX has an optical heliocentric
velocity of $119 \pm 60$ (de Vaucouleurs et al. 1991) and is 
offset spatially from the centroid of the \HI\ concentration. 
Therefore, Ho IX may not be related to either the \HI\ concentration or
the nebula (Fig.\ 1a).  

What might be the origin of the nebula around X-9?
At the distance to M81, the nebula has an \ha\ luminosity 
$L_{{\rm H}\alpha} \sim 1 \times 10^{38} {\rm~ergs~s^{-1}}$ 
(Miller \& Hodge 1994), which is about three times more luminous than the 
bright SNR N49 in the LMC (Vancura et al. 1992).
But the most distinct difference between the two 
is their sizes: $\sim 250$ pc $\times$ 475 pc for the 
X-9 nebula (Miller 1995) vs. 8 pc radius for N49.
If the nebula is heated primarily by a shock (Miller 1995), 
a significant fraction (parameterized here as
$\xi$) of $L_{{\rm H}\alpha}$ may then be produced by the 
excitation as pre-shock hydrogen atoms drift into the post-shock region
(e.g., Cox \& Raymond 1985). Optical 
spectroscopy so far, however, places only an upper limit of $300 
{\rm~km~s^{-1}}$ on the 
shock velocity of the X-9 nebula  (Miller 1995).
Assuming a shock velocity $v_2 \gtrsim 2$ 
(in units of $10^2 {\rm~km~s^{-1}}$ and 
the standard case B (that is, optically thick for all Lyman recombination 
lines), we can estimate from 
$L_{{\rm H}\alpha}$ the pre-shock neutral gas density as 
$n_o \sim (10 {\rm~cm^{-3}}) v^{-1}_2 R_2^{-2}\xi$, 
where $R_2$ (in units of $10^2 
{\rm~pc}$) is the characteristic radius of the nebula. 
The kinetic energy is then $E_k \sim (1 \times 10^{53} {\rm~ergs})
v_2 R_2\xi$ if $n_o$ is spatially uniform. This is likely to be an 
overestimate because the bulk of the H$\alpha$-emitting gas is expected
to arise in dense filaments and clouds in which shock velocity tends to be low.
Furthermore, a substantial fraction of the H$\alpha$ emission may also
be due to the ionization by the X-ray source (\S 4.3). Nevertheless, 
comparisons with known SNRs suggest that the 
X-9 nebula appears to energetic to be due to a single normal supernova (SN).  

Could the optical nebula then be a superbubble produced by multiple 
SNe and fast stellar winds of massive stars
(e.g., Mac Low \& McCray 1988)? 
At the position of X-9, there is indeed a fuzzy blue  object
which may represent a stellar cluster ($m_{\rm B} = 20$; Henkel et al. 1996; 
Miller 1995). If this is the case, the stellar cluster should then have an age 
$t_s \gtrsim 10^7$ yr, as young stars do not appear to
contribute much to the ionization of the optical nebula  and 
no far-UV emission peak appears at the position of X-9 (e.g., Fig.\ 1). 
Comparing $t_s$ with the expansion age of the nebula
$t_e \sim 3R/5v = (1 \times 10^6 {\rm~yr}) 
R_2 v_2^{-1}$, we find  $v \lesssim (5 {\rm~km~s^{-1}})
R_2$. This is less than the turbulence velocity of the ISM 
and too small for shock-heating to effectively produce significant 
\ha\ emission. Therefore, unless being accelerated recently by an SN (or
a hypernova; see later discussion) the optical nebula is probably not 
a superbubble created by a massive stellar cluster.

\subsection{Association of the Optical Nebula and X-9}
 
Using the log(N)-log(S) function from the \rosat\ All Sky Survey (Hasinger et
al. 1998), we estimate that the probability for a chance projection of an X-ray
source comparable to, or brighter than, X-9 within a circle of 
$\sim 10^{\prime\prime}$ radius is only about $\sim 5 \times 10^{-7}$. 
Therefore, X-9 is most likely associated with the nebula.

We speculate that the optical nebula may be directly related to 
the presence of X-9. One possibility is that they were born together.
The nebula may be an interstellar remnant of a hypernova explosion 
that is substantially more energetic than a normal SN.
Hypernovae have been
postulated as the sources of some $\gamma$-ray bursts observed at 
cosmological distances  (Paczy\'nski 1998; Fryer \& Woosley 1998). 
The proposed mechanism for hypernova explosions 
is the collapse of certain massive stars and/or their mergers with 
compact companions. Such an 
event leads to the formation of a BH and provides an extractable 
energy of $\sim 10^{54}$~ergs (M\'esz\'aros, Rees, \& Wijers 1999). 
Hypernovae may also be responsible for some \HI supershells or holes
observed in the interstellar medium (ISM) of nearby galaxies (Efremov,
Elmegreen, \& Hodge 1998; Loeb \& Perna 1998), as well as 
relatively young energetic shell-like  nebulae (Wang 1999). 
The optical nebula around X-9 could well be such a hypernova remnant.
The X-ray source may represent the resultant BH accreting from the
materials falling back from the explosion. Although a binary may 
survive the explosion, the turning-on of a persistent accretion 
phase is expected to be long after
the explosion remnant has disappeared. The soft X-ray contribution from the 
hypernova remnant may be significant in the PSPC spectrum, whereas the GIS 
spectrum is dominated by the bright accretion disk in a higher state,
which makes the detection of the soft component difficult.

Another plausible scenario is that the nebula is currently 
powered by an intense outflow or a wind from X-9 as an accreting X-ray binary. If the nebula is purely caused by this outflow, 
the required mean energy output over the nebula's expansion age
$t_e$ is then a few times $ 10^{39} {\rm~ergs~s^{-1}}$, which is comparable to 
the X-ray luminosity of the source. In principle, the power of 
such an outflow could even be greater than the radiation 
luminosity of X-9, if it is similar to other types of accreting 
BH systems (i.e., Galactic micro-quasars or AGNs). In this case, the IMBH may be a
Population III remnant (e.g., Madau \& Rees 2001) and its companion may be
formed from the collapse of surrounding molecular gas inside 
Concentration I 
(Henkel et al. 1993; Yun et al. 1994; Brouillet et al. 1992; Fig.\ 1).

X-9 may also contribute to the ionization of the nebula. The MCD model of X-9
predicts an integrated ionizing photon rate of $\sim 4 \times 10^{40}
{\rm~s^{-1}}$ over the 0.016--0.2 keV range; photons at higher energies should
mostly escape from the nebula. This rate is about a factor of 10
greater than that estimated for LMC X-1 (a stellar mass BH candidate),
which is surrounded by an X-ray-ionizing nebula (Pakull \& Angebault 1986).
The nebula around X-9 shows indications of 
X-ray ionization. First, the H$\beta$ to H$\alpha$ flux ratio  ($\sim 0.6$)
of the nebula is high, indicating a high electron temperature of $\sim 10^5$~K
(Miller 1995).  Second, the shell-like morphology of the nebula appears
to be rather diffuse, as is expected from the relative long absorption
path-length of soft X-rays and its strong energy dependence 
(e.g., Rappaport et al. 1994). 

Further scrutiny of the above scenarios and their relative importance 
is both desirable and possible. The nature of the blue object is yet to be 
determined: is it the optical counterpart of the accreting system or the 
outflow? 
The outflow may also be probed by observing its nonthermal radio emission,
which should show a persistent flat or inverted radio spectrum 
(e.g., Fender 2001). The 20 cm radio continuum map of Bash \& Kaufman (1986), 
which is centered on M81, does show a 3$\sigma$ contour 
at the position of X-9.
The total flux of $\sim 1$ mJy, if indeed associated with the outflow, 
is substantially greater than that observed from Galactic micro-quasar-like objects.
Detailed optical spectroscopy of the nebula will be especially important 
for determining the expanding velocity of the nebula and will provide 
important constraints on the extreme UV to 
soft X-ray radiation properties of X-9. Such radio/optical 
observations, combined with future high resolution X-ray imaging and spectroscopy, 
should allow for a firm determination of the relation
between the nebula and X-9. 

	In summary, the association of X-9 with 
both the blue object and the shell-like $H\alpha$ nebula as well as 
Concentration I provides an excellent opportunity for studying the nature
of ultraluminous X-ray sources and for characterizing their radiation and outflow as well as their effects on the interstellar medium.

\acknowledgements
The author thanks B. W. Miller and M. Yun for the H$\alpha$ and 
\HI\ images used here, A. D'Onofrio for her help in analyzing the 
archival \asca\ GIS data mentioned in the text, and the referee for
useful comments. The author also appreciates 
comments from S. Immler, R. Williams, K. Wu, and S. N. Zhang on the work,
which was funded by NASA under the grants NAG5--9429 and NAG5--8999.

\end{document}